# Defining a Strategic Action Plan for AI in Higher Education[1]

Nikolaos Avouris
University of Patras, Patras, Greece
avouris@upatras.gr

**Abstract**
This paper discusses key challenges of Artificial Intelligence in Education, with main focus on higher education institutions. We start with reviewing normative actions of international organizations and concerns expressed about the current technical landscape. Then we proceed with proposing a framework that comprises five key dimensions relating to the main challenges relating to AI in higher education institutions, followed by five key strategic actions that the main stakeholders need to take in order to address the current developments. We map these actions to the main stakeholders of higher education and propose a deployment plan. This defines a framework along the dimensions: Challenges, Actions, Stakeholders, Deployment CASD. Examples of AI specific actions at the institutional and individual course level are also provided and discussed.

**Introduction**
We live in accelerated transformative times in education, as AI is increasingly embedded across all levels and forms of formal and informal learning, like in most aspects of life. Common educational applications of AI include personalized learning, AI tutors, automated assessment and feedback, curriculum design support, support for lifelong learning pathways, management of admissions/ progression decisions and many more. As Kasneci et al. (2023) argue, while large language models (LLMs) offer unprecedented support for personalized learning, feedback, and access to knowledge, they also pose complex challenges for academic integrity, transparency, and educational equity. The common practice in most higher education institutions is depicted in the recent Educause Horizon Report (Robert et al., 2025) which confirms the continuous rise of use of AI in universities and outlines the current practice: On one hand, the students are increasingly reported using AI tools like ChatGPT to generate essays, do research, or get help with assignments, while teachers are experimenting with AI-driven lesson planning and assessment. On the other hand, the key stakeholders express serious concerns, about academic integrity, and the erosion of original thought, while teachers express questions about how to teach critical thinking and writing when AI can play that role effectivelly. In the associated faculty survey (Robert et al., 2025), the responses showed that some instructors embrace AI as a learning assistant, guiding students in how to use it ethically, while others design AI-resistant assignments, like oral exams or handwritten essays. Finally, in terms of institutional policies, many universities and schools are rethinking assessment policies and academic integrity rules, while special emphasis is shifting toward AI literacy, that aims at educators helping students learn how to collaborate with AI, not just avoid it. Studies like Vargas-Murillo et al. (2023) literature review of GenAI in higher education, highlight that AI is already reshaping academic practices internationally — prompting institutions to develop policies that balance innovation, fairness, and integrity in AI-assisted learning environments.

There have been various attempts and guidelines for policies development, both from international institutions, like UNESCO, EU, etc, and universities associations, like the APRU maturity framework (Liu & Bates, 2025), that reflects the institutional culture, policies, access,

[1] **Acknowledgement:** This research is co-funded by the Erasmus+ programme of the European Commission (Project "Homo Digitalis", Project number: 101129182)

familiarity and trust to AI, that determine how to design, implement, and evaluate an AI strategy at institutional level.

In this paper, we review the current literature and proposals by international institutions and researchers to define the main challenges related to AI in Education, next we present, given these challenges, key dimensions of concern followed by a set of key strategic actions to address these concerns and deployment of these actions. Next, we define the framework dimensions, starting from the key challenges of AI. The main components of the proposed framework are shown in Figure 1, which, illustrates how challenges inform strategic actions that are mapped to stakeholders and implemented through deployment strategies.

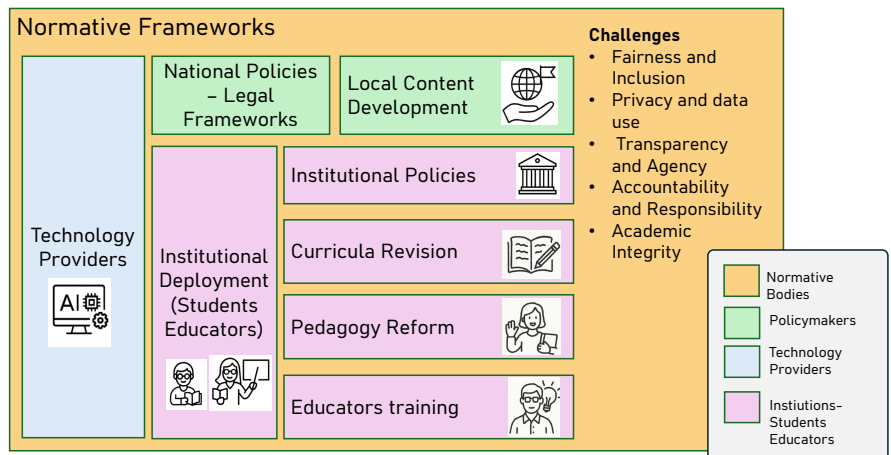

Figure 1. Components of the CASD (Challenges, Actions, Stakeholders, Deployment) framework

## Normative foundations of AI

We start with identifying the main challenges related to AI. In order to do that, we first review normative frameworks that have been issued by authoritative bodies, concerning design, deployment and governance of AI. A key reference is the widely adopted framework proposed by Unesco (2021), that is grounded in human rights, inclusion, sustainability, and peace. In particular, this framework outlines four foundational values that may be affected by AI: (i) *Human Rights and Human Dignity*, i.e. AI must not infringe on learners' fundamental rights (e.g., privacy, autonomy, expression). (ii) *Environmental Sustainability* that commands that AI systems should be designed and used in ways that support sustainability, especially relevant for educational technology systems consuming large-scale data and energy, (iii) *Diversity and Inclusiveness*, AI must be inclusive, respecting different languages, cultures, identities, from which stems the need for localization of AI tools to reflect educational and linguistic diversity, and finally (iv) *Living in Peaceful, and Just Societies*, which implies that education should prepare learners to engage critically and ethically with AI in civic life. A value that emphasizes digital literacy, fairness, and social cohesion. Next the Unesco framework proceeds with proposal of ten key principles, that are actionable ethical rules, based on these fundamental values. The principles, with examples of their implication in the education sector, are:

1. *Privacy and Data Protection*, which relates to possible mass collection of student data via e-proctoring or educational technology platforms

2. *Fairness and Non-Discrimination*, which concerns algorithmic bias disadvantaging underrepresented student groups, and marginalized languages

3. *Proportionality and Do No Harm*, resulting in the principle that AI systems should not cause more harm or risk than necessary to achieve their intended purpose.

4. *Safety and Security*, relating to exposure of minors to manipulative or unsafe systems

5. *Human Oversight and Determination*, that concerns the danger of teachers loosing control over AI tutoring systems
6. *Accountability and Responsibility*, that relates to cases of unclear responsibility for automated decisions in learning analytics
7. *Transparency and Explainability*, concerning black-box algorithms in grading or recommendation systems
8. *Awareness and Literacy*, e.g. cases of students and staff to be unaware of how AI tools work
9. *Sustainability*, that relates to overuse of high-resource AI tools that burden IT infrastructure
10. *Multi-stakeholder Governance*, that proposes that governance of AI must include diverse stakeholders: educators, students, parents, technologists, policymakers, civil society.

These ten principles consolidate ideas found in many other, similar frameworks or guidelines, like the Unesco Guidelines for GenAI in education and research (Unesco, 2023), the Council of Europe Report on AI and Education (Holmes et al. 2022), the European Commission Guidelines, a practitioner-oriented resource that aim at promoting the responsible integration of AI and data in education, (European Commission, 2022), and the Unesco report ( Miao et al., 2021) with guidelines for policy makers, that translates high-level ethical and strategic frameworks into practical policy considerations specifically for education systems.

## The AI Challenges

Next, summarizing the proposals of the discussed normative frameworks, we define the five challenges that need to be addressed making *Dimension C* of our framework:

Challenge C1: *Fairness and Inclusion*. Artificial intelligence can unintentionally reinforce bias. Whether in grading, tutoring, or content recommendations, AI may reflect existing social inequalities. It can also exclude students whose needs or languages are underrepresented in the data. To ensure AI in education is fair, we must proactively design for inclusion—considering accessibility, multilingual support, and cultural diversity. Fairness isn't just a technical issue; it's an ethical commitment to serve all learners.

Challenge C2: *Privacy and data use.* AI systems rely heavily on data—often personal, often sensitive. In education, this includes student behaviour, assessment patterns, even emotional states. Without robust protections, AI tools can compromise privacy and data sovereignty. We must ensure that students and teachers are not merely data sources, but informed participants.

Challenge C3: *Transparency and Agency.* It is known that many AI systems operate as black boxes. They offer recommendations or assessments, but their reasoning is obscure—even to the educators using them. This undermines trust and can erode the autonomy of both students and teachers. Transparency is essential: educators should understand how tools work, and students should retain control over their learning. AI must support—not replace—human judgment and critical thinking.

Challenge C4: *Accountability and Responsibility.* When the AI system misjudges a student's answer or gives biased feedback, the question raises who is responsible? Without clear accountability, these errors go unchecked. Institutions must ensure that AI use in education is supervised, auditable, and ultimately human-led. Ethical use requires human oversight, clear escalation paths, and shared responsibility between developers, educators, and administrators.

Challenge C5: *Academic Integrity*. This challenge is not explicitly present in the normative frameworks, yet it is the main concern of stakeholders, as discussed in Robert et al., (2025), as it touches on many ethical principles of AI. Current AI tools like Large Language Models chatbots pose new questions about originality, authorship, and fairness in education. We need to rethink integrity, not just as enforcement, but as empowerment: teaching students to use AI responsibly, reflecting on how it's used, and designing assignments that reward genuine thinking.

In the next section we proceed with proposing a set of strategic actions to address the discussed here challenges.

**The Strategic Actions**
Many proposals have been made for actions to be taken to address the challenges of AI in education. For instance, European Commission (2022) and Unesco (2021b) contain such ideas. Chan (2023) and Chan & Colloton (2024) have made comprehensive proposals for a policy education framework for university teaching and learning. At the level of institutional policies, there have been many publications reviewing specific countries or compare policies across areas. The review of Dell'Erba et al. (2025) concerns the policy documents of top-ranked universities, focusing on the innovation-regulation tension, while Jin et al. (2025) analyse adoption of GenAI policies in 40 universities across six global regions focusing on the characteristics of the institutions, like compatibility, trialability, observability, and the communication roles/responsibilities, using innovation diffusion theoretical framework. Finally studies in specific countries reveal the characteristics of their educational systems. Abir & Zhou (2025) who studied Japanese universities, found out their cautious stance toward GenAI, emphasizing concerns about academic integrity, transparency, and unintended misuse, while Li et al. (2025) performed a cross-national comparative study involving over 100 policy documents across the US, Japan and China. Using these findings they developed the UPDF-GAI model for guiding universities in developing AI policies. It is interesting to observe that from this study, the different orientation of different academic systems emerged, with the US leaning toward faculty autonomy, Japan emphasizing government-regulated frameworks, and China generally aligned with a centralized, top-down model, prioritizing AI integration and technology-driven implementation over early-stage policy structuring. Based on these ideas, we provide here a set of key strategic actions, defining the *Action Dimension* of our framework.

*Action 1*: *Revise Curriculum.* We need to make drastic changes to the curricula content, in order to teach learners how to critically analyse information and verify their sources and develop digital literacy skills that can help them better understand how AI technologies work and how to use them responsibly, this needs to be adapted to the requirements of each subject matter.

*Action 2*: *Reform Pedagogy.* The way we teach should be adapted in the times of AI. Emphasis should be given on assignments and assessments conducted in class, so that the use of AI for solving problems and completing assignments can be monitored and the unfair use of AI can be avoided.

*Action 3*: *Training of Educators.* We need to design training programs for educators so that they can better understand AI technologies and effectively integrate their capabilities into the educational process.

*Action 4*: *Support Local Content.* We need to ensure that the AI models used in education can be properly trained in the local language and with local data, by providing high-quality, machine-readable local content. An interesting relevant observation is that the corpora used for training AI models, have often bias in certain languages and cultures. For instance, the OSCAR23[2] corpus, a large-scale, multilingual text dataset, with around 1 trillion words, derived from the Common Crawl web archive, used for training many AI models, contains 48% of content in English, and just 0.71% in Greek, or 0.04% in Albanian, while some minority languages are not existent.

*Action 5: Establish Policy and Governance.* Educational institutions must establish clear regulations regarding the use of AI technologies, defining what is permitted and what is not, as well as the consequences of related violations of academic integrity.

It is obvious from the list of these strategic actions that these responses require participation of different stakeholders of the higher education ecosystem. In the next section we will discuss the role of the five key stakeholders in these actions.

---
[2] https://oscar-project.github.io/documentation/versions/oscar-2301/

## The Roles of Stakeholders

We consider the following five key stakeholders of the higher education ecosystem: the Students, Educators, Institutions, Policymakers and Technology providers. We will briefly describe next their roles in the proposed strategic actions.

*S1. Students*. These are the main stakeholders as they are the target of the main mission of the higher education system. They participate as receivers of the new services. They are affected by *Revised Curriculum*, as through this action's results they gain AI literacy and awareness of ethical use, by *Reformed Pedagogy*, as they experience new ways of teaching, for instance, more inclusive, AI-aware assessments and by *Support Local Content*, as they have, as a result, access to AI tools relevant to their language and culture.

*S2. Educators*. The faculty of higher education need to adapt to the new conditions, participating actively in all proposed strategic actions, In *Revised Curriculum*, they are asked to teach new topics encouraging critical thinking and AI literacy. For *Reform Pedagogy*, they need to adapt their teaching using new AI-resilient assessment, In *Train Educators*, to participate in faculty training actions to acquire competence to use and explain AI tools, in *Support Local Content* to teach using culturally appropriate and multilingual tools, and in *Policy & Governance,* to implement the AI policies of the institutions.

*S3. Institutions*. The higher education institutions are asked to play a key role according to the proposed framework. For *Revise Curriculum* They need to develop curricula that align with digital competencies and AI use, for *Reform Pedagogy,* they need to encourage pedagogy that includes fairer evaluation methods, for *Train Educators,* to provide training programs and support systems for their faculty, in *Support Local Content*, to invest in open educational resources and local repositories, and in *Policy & Governance*. to establish oversight and internal governance structures.

*S4. Policy Makers*. These are institutions at the national level (e.g. Ministries of Education) or international level (e.g. EU, Unesco, etc.). There are already recommendations for policies, as discussed in previous section, while the EU in its AI Act makes explicit reference to the education sector as a high-risk domain with obligations for technology providers and for institutions. However, there are not explicit policies in most parts of the world. The role of the policy making bodies is crucial. In *Revise Curriculum,* the policies should contain curriculum guidelines that include AI ethics and AI literacy, for *Reform Pedagogy*, policies should exist that encourage compliance with national and international standards for ethical AI based pedagogy. For *Train Educators*, there should be funding and policies that mandate teacher training on AI, for *Support Local Content*, policies that ensure content diversity and inclusion, with local language and culture. *Policy & Governance,* there should regulate all aspects of development and use of AI in education, based on ethical frameworks.

*S5. Technology Providers*. The responsibilities of the AI technology providers are high according to the normative recommendations and the existing legal framework, e.g. the EU AI Act, foresee that the developers of the technology bear most of the legal burden relating to compliance, documentation and risk management. In the frame of *Reform Pedagogy,* they are required to develop technologies supporting ethical and creative pedagogy, for *Support Local Content,* they are required to localize the developed AI systems for underrepresented languages or contexts, and for *Policy & Governance*, they need to take all necessary actions to comply with standards; ensure ethical, transparent, equitable technologies.

Having outlined the strategic actions and their mapping to key stakeholders, we now turn to the question of how this framework can be implemented effectively in institutional settings.

## Framework Deployment

As discussed in the previous sections, the proposed here framework requires the active involvement of all key stakeholders of the higher education ecosystem. Given that the technological ground of AI is in continuous evolution, a deployment of such framework needs on one hand, to be flexible to accommodate this shifting technological background, and on the other hand, to match the readiness and the culture of the specific institution or educational system. Liu & Bates (2025) identify five important dimensions, that make up the CRAFT framework for AI strategy implementation in a higher education institution: Culture, Rules, Access, Familiarity, Trust. These dimensions relate to alignment of the AI strategy with institutional mission, pedagogical values, and change-readiness (the *Culture* dimension), to definition of rules that provide the governance layer: ethical guidelines, acceptable use policies, and accountability structures, to ensure equitable access to AI tools to training, and infrastructure across all stakeholders (the *Rules* dimension), to build AI literacy, technical skills, as well as critical, ethical, and creative competences (the *Familiarity* dimension), and finally the *Trust* dimension, that facilitates transparency, dialogue, and confidence in institutional strategy and tools. This leads to a roadmap for deployment of AI policies, following a sequence of steps: (i) Start with Culture, by surveying faculty/ student values, identify champions, and establish how ready the institution is for establishing these policies, (ii) develop the Rules through outlined actions, like co-creation of policies on ethical use of AI, especially in assessment (guidelines for pedagogy reform and new curricula), (iii) next, there is need to ensure access, this is done by deploying institutionally supported AI tools, while special measures have to be providing in ensuring language/inclusive access, (iv) then build familiarity, i.e. to offer cross-disciplinary AI literacy programs; empower staff (teacher training action), and (v) finally, in order to establish trust, there is need to share data about AI use outcomes; and invite ongoing community input. Furthermore, Chan & Colloton (2024), provide guidelines to enhance the deployment framework: As a starting point, they focus on changing institutional culture, from awareness of AI's educational implications to proactive leadership engagement and identification of faculty champions, role-based responsibilities, feedback mechanisms, and aligning AI strategy with institutional mission. They also provide a policy blueprint for institutional AI guidelines, including templates for acceptable use, ethical standards, and operational processes, while they advocate an iterative, inclusive design process with experts, faculties, librarians, disability services, and IT teams participating in co-design.

## Conclusions and Future Directions

In this paper we have reviewed normative approaches to AI in education and outlined a framework for design and deployment of policies for higher education. Concluding this review, we provide some references to resources that can help with implementing the proposed action plan. First at the level of individual courses, there are various examples. Stanford Teaching Commons (2023) provide templates to instructors through an AI Teaching guide, with specific guidelines on course AI policies, AI-resistant assignments, and support for AI literacy. Avouris et al. (2025) have described the experience of re-design of a computer science course in order to be AI-resilient, providing examples on re-design of assessment strategies, pedagogy and learning objectives of the course, while Chan & Colloton (2024) propose strategies for redesigning assignments and evaluation, the *Six assessment redesign pivotal strategies*.

In conclusion, as AI technologies continue to evolve, higher education institutions must adopt dynamic, inclusive, and ethically grounded strategies. The proposed framework offers a foundation for such action, yet ongoing collaboration between educators, students, policymakers, and developers remains essential. We should however consider the risks and barriers of the strategic framework, like institutional resistance, funding, faculty disposition, that may affect its implementation. Future work should focus on evaluating institutional readiness, piloting policy innovations, and fostering global exchange of good practices in AI-enhanced education. Frameworks like the DigiReady+ (Chounta et al. 2024), that measure digital readiness of higher education institutions can be useful tools in supporting this process.

# References

Abir, M. G. H., & Zhou, K. Z. (2025). Examining Generative AI Policies in Japanese Universities: A Qualitative Perspective. Available at SSRN 5374222.

Avouris N., Sgarbas K., Caridakis G., Sintoris C., (2025). Teaching Introduction to Programming in the times of AI: A case study of a course redesign, *12th Panhellenic Conference on Computer Science Education*, Rhodes, October, 2025.

Chan, C. K. Y. (2023). A comprehensive AI policy education framework for university teaching and learning. *International journal of educational technology in higher education*, *20*(1), 38.

Chan, C. K. Y., & Colloton, T. (2024). *Generative AI in higher education: The ChatGPT effect*. Taylor & Francis.

Chounta, I. A., Ortega-Arranz, A., Daskalaki, S., Dimitriadis, Y., & Avouris, N. (2024). Toward a data-informed framework for the assessment of digital readiness of higher education institutions. *International Journal of Educational Technology in Higher Education*, 21(1), 59.

Dell'Erba, C., Ruffini, L., Silva, R., & Consoli, L. (2025). Mapping global generative artificial intelligence guidelines in higher education: the ambiguous balance between innovation and regulation. In *EduLearn25 Proceedings* (pp. 5710-5720). IATED.

Directorate-General for Education, Y. (2022). Ethical guidelines on the use of artificial intelligence (AI) and data in teaching and learning for educators. *Publications Office of the European Union*. https://data.europa.eu/doi/10.2766/153756

European Commission, (2022). Ethical guidelines on the use of artificial intelligence (AI) and data in teaching and learning for educators, *Publications Office of the European Union*. https://data.europa.eu/doi/10.2766/153756

Jin, Y., Yan, L., Echeverria, V., Gašević, D., & Martinez-Maldonado, R. (2025). Generative AI in higher education: A global perspective of institutional adoption policies and guidelines. *Computers and Education: Artificial Intelligence*, 8, 100348.

Kasneci, E., Sessler, K., Küchemann, S., Bannert, M., Dementieva, D., Fischer, F., Gasser, U., Groh, G., Günnemann, S., Hüllermeier, E., Krusche, S., Kutyniok, G., Michaeli, T., Nerdel, C., Pfeffer, J., Poquet, O., … Kasneci, G. (2023). ChatGPT for good? On opportunities and challenges of large language models for education. *Learning and Individual Differences*, 103, 102274.

Li, M., Xie, Q., Enkhtur, A., Meng, S., Chen, L., Yamamoto, B. A., ... & Murakami, M. (2025). A Framework for Developing University Policies on Generative AI Governance: A Cross-national Comparative Study. *arXiv preprint* arXiv:2504.02636.

Liu, D. Y. T., & Bates, S. (2025, January). Generative AI in Higher Education: Current Practices and Ways Forward. *Association of Pacific Rim Universities Report*, available from https://biblioteca.unisced.edu.mz/

Miao, F., Holmes, W., Huang R. Zhang H. (2021). Artificial Intelligence and Education. Guidance for Policy-makers. *United Nations Educational, Scientific and Cultural Organization (UNESCO)*, Paris, France.

Robert, J., Muscanell, N., McCormack, M., Pelletier, K., Arnold, K., Arbino, N., Young, K. & Reeves, J. (2025): *2025 EDUCAUSE Horizon Report*, Teaching and Learning Edition

Stanford Teaching Commons (2023) *Artificial Intelligence Teaching Guide*, available from https://teachingcommons.stanford.edu

UNESCO (2023). Guidance for Generative AI in Education and Research, *United Nations Educational, Scientific and Cultural Organization (UNESCO)*, Paris.

UNESCO, C. (2021). Recommendation on the ethics of artificial intelligence. In *United Nations Educational, Scientific and Cultural Organization (UNESCO)* Paris.

Vargas-Murillo, A. R., Pari-Bedoya, I. N. M. de la A., & Guevara-Soto, F. de J. us. (2023). Challenges and Opportunities of AI-Assisted Learning A Systematic Literature Review on the Impact of ChatGPT Usage in Higher Education. *International Journal of Learning, Teaching and Educational Research*, 22(7), 122–135.